\documentclass[apj, revtex4]{emulateapj}

\usepackage{amsmath}
\usepackage[T1]{fontenc}
\usepackage{float}

\usepackage{booktabs}
\usepackage{longtable}
\usepackage{epsfig,natbib} 
\usepackage{enumerate}
\usepackage{multirow}
\usepackage{subfigure}
\usepackage{gensymb}
\usepackage{morefloats}
\usepackage{lscape}
\usepackage{graphicx} 
\usepackage{color}

\shorttitle{Stellar \& Planetary Parameters for Late Type Dwarf Systems} 
\shortauthors{Martinez et al.} 

\begin{document}

\title{Stellar \& Planetary Parameters for {\it K2}'s Late Type Dwarf Systems from C1 to C5}

\author{Arturo O. Martinez\altaffilmark{1,2,3},
Ian J.~M.~Crossfield\altaffilmark{4,5,6}, 
Joshua E. Schlieder\altaffilmark{7,8,9}, 
Courtney D. Dressing\altaffilmark{10,6},
Christian Obermeier\altaffilmark{11,12}, 
John Livingston\altaffilmark{13}, 
Simona Ciceri\altaffilmark{14}, 
Sarah Peacock\altaffilmark{4}, 
Charles A. Beichman\altaffilmark{8},
S\'{e}bastien L\'{e}pine\altaffilmark{2}, 
Kimberly M. Aller\altaffilmark{15}, 
Quadry A. Chance\altaffilmark{16},
Erik A. Petigura\altaffilmark{13,17}
Andrew W. Howard\altaffilmark{18}, 
Michael W. Werner\altaffilmark{19}
}

\altaffiltext{1}{Department of Astronomy, San Diego State University, 5500 Campanile Drive, San Diego, CA, USA}
\altaffiltext{2}{Department of Physics and Astronomy, Georgia State University, GA, USA}
\altaffiltext{3}{Visiting Researcher, Steward Observatory, University of Arizona, Tucson, AZ, USA} 
\altaffiltext{4}{Lunar \& Planetary Laboratory, University of Arizona, Tucson, AZ, USA}
\altaffiltext{5}{Astronomy and Astrophysics Department, UC Santa Cruz, Santa Cruz, CA, USA}
\altaffiltext{6}{NASA Sagan Fellow}
\altaffiltext{7}{NASA Ames Research Center, Moffett Field, CA, USA}
\altaffiltext{8}{NASA Exoplanet Science Institute, California Institute of Technology, Pasadena, CA, USA}
\altaffiltext{9}{NASA Postdoctoral Program Fellow}
\altaffiltext{10}{Division of Geological and Planetary Sciences, California Institute of Technology, Pasadena, CA, USA}
\altaffiltext{11}{Max Planck Institut f\"{u}r Astronomie, Heidelberg, Germany}
\altaffiltext{12}{Max-Planck-Institute for Extraterrestrial Physics, Garching, Germany}
\altaffiltext{13}{Department of Astronomy, Graduate School of Science, The University of Tokyo, 7-3-1 Bunkyo-ku, Tokyo 113- 0033, Japan}
\altaffiltext{14}{Department of Astronomy, Stockholm University, SE-106 91 Stockholm, Sweden}
\altaffiltext{15}{Institute for Astronomy, University of Hawai'i at M\={a}noa, Honolulu, HI, USA}
\altaffiltext{16}{Steward Observatory, University of Arizona, Tucson, AZ, USA}
\altaffiltext{17}{Hubble Fellow}
\altaffiltext{18}{Department of Astronomy, California Institute of Technology, Pasadena, CA 91125, USA}
\altaffiltext{19}{Jet Propulsion Laboratory, Pasadena, CA, USA}

\begin{abstract}

The NASA {\it K2} mission uses photometry to find planets transiting stars of various types. M dwarfs are of high interest since they host more short-period planets than any other type of main-sequence stars and transiting planets around M dwarfs have deeper transits compared to other main-sequence stars. In this paper, we present stellar parameters from K and M dwarfs hosting transiting planet candidates discovered by our team. Using the SOFI spectrograph on the European Southern Observatory's New Technology Telescope, we obtained R $\approx$ 1000 $J$-, $H$-, and $K$-band (0.95 - 2.52 $\micron$) spectra of 34 late-type {\it K2} planet and candidate planet host systems and 12 bright K4-M5 dwarfs with interferometrically measured radii and effective temperatures. Out of our 34 late-type {\it K2} targets, we identify 27 of these stars as M dwarfs. We measure equivalent widths of spectral features, derive calibration relations using stars with interferometric measurements, and estimate stellar radii, effective temperatures, masses, and luminosities for the {\it K2} planet hosts. Our calibrations provide radii and temperatures with median uncertainties of 0.059 R$_{\odot}$ (16.09\%) and 160 K (4.33\%), respectively. We then reassess the radii and equilibrium temperatures of known and candidate planets based on our spectroscopically derived stellar parameters. Since a planet's radius and equilibrium temperature depend on the parameters of its host star, our study provides more precise planetary parameters for planets and candidates orbiting late-type stars observed with {\it K2}. We find a median planet radius and an equilibrium temperature of approximately 3$R_{\oplus}$ and 500 K, respectively, with several systems (K2-18b and K2-72e) receiving near-Earth-like levels of incident irradiation.

\end{abstract}

\subjectheadings{methods: data analysis -- stars: fundamental parameters -- stars: late-type -- planetary systems -- techniques: spectroscopic}

\section{Introduction}
\label{sec:introduction}
Small, low-luminosity M dwarfs are the most common type of star in the Galaxy, but their properties are less well understood than those of hotter solar-type stars. There are still significant discrepancies between theoretical models and observations of M dwarf spectra \cite[e.g.][]{hoeijmakers:2015}, and we are still uncertain as to why the occurrence rate of small, short-period planets is higher for M dwarfs and the occurrence rate of a gas giants (on both close and wide orbits) is lower for M dwarfs when compared to solar-like stars, as shown in studies of the {\it Kepler} field \citep{dressing:2013,gaidos:2014b,morton_swift:2014,dressing:2015,muirhead:2015} and other surveys \citep{shields:2016}. There are a few exceptions to the low occurrence rate of gas giants around M dwarfs; there has been at least one confirmed gas giant orbiting an M dwarf \citep{johnson:2012}.

Fortunately, the discovery of exoplanets around M dwarfs is much easier when compared to finding exoplanets around Sun-like stars. For example, while a transiting 2$R_{\oplus}$ planet would have a transit depth of 0.03\% when orbiting the Sun, that same planet would have a transit depth of 0.5\% for an M5 dwarf. Using planet candidates from the original {\it Kepler} mission, \cite{howard:2012} and \cite{mulders:2015a,mulders:2015b} showed that the occurrence rates of small planets are higher for M dwarf than for any other main-sequence star. Other surveys, such as MEarth \citep{charbonneau:2009,bertacharbonneau:2015} and Transiting Planets and Planetesimals Small Telescope, have also successfully identified interesting new planets transiting M dwarfs \citep{gillon:2016}. Additionally, M dwarfs provide our best chances to identify nearby potentially habitable planets since the habitable zone around M dwarfs, when compared to those around other main-sequence stars, is closer to the M dwarf due to the its lower luminosity. This is exemplified by the discovery of Proxima Centauri b, a small, likely temperate planet orbiting the closest star to the Sun \citep{anglada-escude:2016,damasso:2016}.

Host star properties must be well understood in order to be able to derive planet properties. Unfortunately, the stellar properties of M dwarfs are challenging to predict from photometry \cite[due to M dwarfs being intrinsically faint and the modeling uncertainties as described above and by][]{mann:2015}. The most accurate parameters of M dwarfs are derived from interferometric data \citep{boyajian:2012b} or photometric and spectroscopic observations of double-lined eclipsing binaries \citep{torres:2010}.

For systems where such observations are not feasible, several authors have developed a calibration method based on medium-resolution, near-infrared spectra in order to infer the stellar properties of these M dwarfs from empirical observations \citep{mann:2015,newton:2015,terrien:2015} and stellar models \citep{rojas-ayala:2012}, while others have applied similar empirical calibration techniques to the optical part of the spectrum \citep{neves:2014,maldonado:2015}. By measuring the equivalent widths (EWs), or the strength of any given absorption feature one can calculate stellar parameters by calibrating from a reference sample with previously measured parameters of interest. Since the EW of an absorption feature varies with photospheric temperature and surface gravity, this approach allows these parameters (and related quantities, like stellar radius and mass) to be calculated. 

Using the repurposed {\it Kepler} spacecraft, the {\it K2} mission is continuing to observe many stars in the Galaxy in the search for more exoplanets \citep{howell:2014}. However, {\it K2} has some limitations. With just two (out of four) operating reaction wheels, the spacecraft can observe only along the ecliptic plane with observation windows of 80 days per campaign.  Nonetheless, {\it K2} has provided astronomers with powerful data enabling a large number of candidate and confirmed exoplanets \citep{vanderburg:2014, crossfield:2015a, foremanmackey:2015, huang:2015, montet:2015, sanchis-ojeda:2015, sinukoff:2015, crossfield:2016}.

In this paper we analyze medium-resolution, near-infrared spectra of candidate planetary systems detected by {\it K2} to provide updated stellar and planetary parameters. We measure EWs to infer stellar radii and effective temperatures, and subsequently planetary radii and equilibrium temperatures. In \S\ref{sec:targ_select}, we briefly explain our target selections and how we compiled our planet candidate list. In \S\ref{sec:observations}, we describe our observational techniques, data reduction, and various calibration samples. In \S\ref{sec:spectral_analysis}, we explain the process by which we obtain our stellar and planetary parameters and compare our derived stellar parameters with those of previously spectroscopically and interferometrically measured stellar parameters. In \S\ref{sec:conc_future}, we summarize our results and describe future work relevant to this paper.

\section{Target Selection and Planet Candidate Search}
\label{sec:targ_select}
We initially selected our {\it K2} M dwarf candidates from Campaigns 1 through 5. Our team selected and proposed late-type dwarf targets to the {\it K2} mission as described by \cite{crossfield:2016}. In brief, we selected targets as being likely low-mass dwarfs by a combined color and proper motion cut with ($V - J$) > 2.5, $V$ + 5 $\log \mu$ + 5 < 10, and (6$V$ - 7$J$ - 3) < 5 $\log \mu$ \cite[where $\mu$ is the proper motion;][]{crossfield:2015a}. The combination of the color and proper motion cut greatly reduces giants from our sample and further narrows down the M dwarf candidate list. Finally, we imposed a magnitude limit of $Kp$ < 16.5 mag \citep{crossfield:2016}.

We further identify likely low-mass planet-hosting dwarf stars, as explained in \cite{crossfield:2016}. In brief, we used the \texttt{TERRA} algorithm \citep{petigura:2013a} to search for planet transits that have a signal-to-noise ratio (S/N) > 12, which are called threshold-crossing events (TCEs). TCEs are required to have orbital periods of $P \geq$ 1 day and to have at least three transits. These restrictions, along with the diagnostic tests that \texttt{TERRA} provides, show whether the object is a candidate transiting planet, binary star system, another variable object, or noise. If a planet candidate is found, \texttt{TERRA} is iteratively repeated after removing the identified transit signals \cite[described by][]{sinukoff:2016} to see whether there are any additional planets in the system.

\section{Observations}
\label{sec:observations}
We acquired our infrared spectra at the 3.58 m European Southern Observatory (ESO) New Technology Telescope (NTT) using the SOFI spectrograph \citep{moorwood:1998} as part of program 194.C-0443 (PI: I.~J.~M.~Crossfield). We observed through 13 full or partial usable nights in 2015 and 2016. We used two grisms, red and blue, to produce a total spectrum for each object spanning a continuous wavelength range from 0.95 to 2.52 $\micron$\footnote{The blue grism spans the wavelength range from 0.95 to 1.64 $\micron$ while the red grism spans the wavelength range from 1.53 to 2.52 $\micron$. Note, there is a small overlap from both grisms in the $H$-band, thus allowing the fully-reduced spectra of all of our stars to be continuous.} at a resolution of {\it R} $\approx$ 1000. Dome flats and lamps were either taken at the start or the end of each observing night. Our observation sample comprises 34 stars observed by {\it K2} in fields 1 through 5, along with 12 bright K and M dwarfs with interferometrically measured stellar parameters (refer to Table \ref{tab:calibration} for our calibration sample). 

For all observations, we used an ABBA nodding pattern to obtain the spectrum of the object, while removing the spectrum of the background, including sky emission lines and dark current. The exposure times for each frame range from the minimum allowed exposure time (1.182 s) to 120 s. We typically took at least six separate spectra (for each grism) for all the targets. Either immediately before or after each M dwarf candidate, we observed a nearby A0V star for telluric corrections. If the observation for one grism took more than 10 minutes, its A0V calibrator would be taken before the start of the first grism and then taken again after the second grism exposure had finished, for their respective grisms. We identified suitable A0V stars using the IRTF's online tool\footnote{\url{http://irtfweb.ifa.hawaii.edu/cgi-bin/spex/find\_a0v.cgi}}.

\subsection{Data Reduction}

The raw data taken at the NTT were reduced by using a combination of Python, Image Reduction and Analysis Facility (IRAF) software,\footnote{Developed at the National Optical Astronomy Observatory} and using various Interactive Data Language (IDL) programs. We flat-fielded the raw spectra in order to correct for any pixel-to-pixel variation. Wavelength calibrations were done by taking Xe arc spectrum for both grisms either at the beginning or at the end of the night. Using IRAF, emission lines from the taken Xe arc frames were manually selected by comparing them to the SOFI manual\footnote{Provided by ESO.}. One-dimensional spectra were then extracted for identifying the star's spectrum. IRAF had difficulty tracing the 2D spectra of our fainter targets, so for these stars we used brighter stars during that night to define a static extraction aperture.

We subsequently used the IDL routines of \cite{vacca:2003} to process our spectra. First, with \texttt {xcombspec} \cite[from the SpeXtool software package by][]{cushing:2004}, we combined multiple exposures for a given grism of an object into one spectrum. Any spectra that are not shown to have similar spectral features with the other exposures for that star and grism were excluded. 

We corrected for telluric absorption by using our A0V spectra with the \texttt {xtellcor\underline{ }general} routine. Spectra of A0V stars were used since these stars are mostly composed of featureless spectra, with the exception of hydrogen absorption. Differences between the hydrogen lines in the A0V and a model Vega spectrum were corrected for, and then the object's spectrum was divided by the resulting telluric spectrum of the A0V; the observations for the telluric calibrator were usually taken within a short time (approximately 15 minutes) and have a similar airmass (within 0.3 airmass) to the object \citep{rojas-ayala:2012}. We note that for some of the observations, the telluric calibrator's spectrum was sufficiently different from that of Vega that some residual H lines remain in the M dwarf candidate's spectrum. Additionally, the large differences in airmass left residual telluric features in some of the spectra, and any spectra that were contaminated were removed from our analysis. 

\begin{figure}[ht!]
\begin{center}
\includegraphics[width=3in]{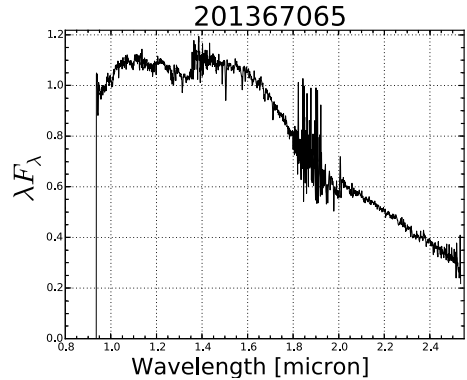}
\caption{\label{fig:k2_3_spectrum} Sample spectrum of one of our {\it K2} targets (EPIC 201367065 or K2-3) that covers a continuous wavelength from 0.95 to 2.52 $\micron$ and is normalized to the median flux value. Note that we ignore regions heavily contaminated by telluric features (e.g., wavelength ranges that are within 1.35-1.45 $\micron$ and 1.80-1.95 $\micron$). After data reduction is complete, we trim an approximate 0.01-0.02 $\micron$ off the edges of the wavelength ranges. Spectra of all our stars are available as electronic supplements to this paper.}
\end{center}
\end{figure}

The last step for the reduction process was to combine the two different grisms using \texttt {xmergexd}. We then used several strong absorption features in each spectrum to correct for radial velocity (RV) shifts and/or offsets in our wavelength calibration. Finally, we interpolated all spectra to put them on the same wavelength scale. All of the objects in our sample have a S/N that ranged from 20 (for the faint {\it K2} targets)\footnote{{\it K2} targets that had a S/N of 20 were removed from the likely low-mass dwarf list, thus making our final 34 star sample.} to over 200 (for the brighter, interferometric calibration targets). We show a representative reduced spectrum in Figure \ref{fig:k2_3_spectrum}.

\subsection{Calibration Sample}

We applied the relations from a variety of works, such as \cite{neves:2014} and \cite{maldonado:2015}, which fit various functions for a variety of EW ratios, and \cite{terrien:2015}, which measured {\it H}-band atomic features, to stars with previously measured radii and/or effective temperatures. However, stars that are interferometrically measured are preferred to these samples since measurements from interferometry are more accurate and precise when compared to spectroscopic, EW-based methods. Although most interferometrically measured stars lie too far north to be observed with SOFI, we managed to obtain spectra of 12 stars with previously interferometrically determined stellar radii and effective temperatures. These stars form our calibration sample, and their properties are summarized in Table \ref{tab:calibration}.

\section{Spectral Analysis}
\label{sec:spectral_analysis}

\cite{mould:1976} was the first to use infrared absorption line strengths to estimate the radii and effective temperatures of low-mass dwarfs. The strengths of absorption features corresponding to a given element or molecule depend on the effective temperatures of the star. Changing the temperature of the star then changes the electronic (or vibrational) population levels of the element (or molecule) in the M dwarf atmosphere. M dwarf radii are related to their effective temperatures so that they roughly follow a linear relation from 4700 K \& 0.7 $R_{\odot}$ down to at least 3300 K \& 0.3 $R_{\odot}$. Some of the absorption features in the spectrum can also present information about the stellar surface gravity. The lines of alkali elements, for example, are affected by surface gravity and can then be used to distinguish old dwarf stars, young dwarf stars, and giants with similar temperatures \citep{spinrad:1962, steele_jameson:1995, lyo:2004, schlieder:2012b}.

The EW is defined by the following equation:
\begin{equation}
{EW_{\lambda} = \int_{\lambda_{1}}^{\lambda_{2}} \! \left[1 - \frac{F(\lambda)}{F_{c}(\lambda)} \right] d\lambda}
\end{equation}
where {\it F}(\(\lambda \)) is the flux of the absorption feature between $\lambda_{1}$ and $\lambda_{2}$, and {\it $F_{c}$}(\(\lambda \)) is the continuum flux. We investigate the features used by \cite{cushing:2005}, \cite{rojas-ayala:2012}, \cite{newton:2014}, and \cite{newton:2015} for our work. The features, shown in Table \ref{Tab:lines}, are slightly adjusted owning to differences in resolution of the spectrographs - typically our integration ranges are slightly wider than those previously presented. Additionally, any spectral line doublets and molecular bands used in our empirical indices are treated as single features in the the EW calculations. The blue continuum and red continuum of each feature are also adjusted such that they would not overlap with any nearby feature windows. In the following sections, we describe the steps that are taken to infer the stellar and planetary parameters using these EW measurements of our {\it K2} and calibration samples.

\subsection{Spectral Classification}
\label{SpT_class}
We visually estimated the spectral types (SpT) of each of our stars by comparing our SOFI spectra to spectra of standard stars in the IRTF Spectral Library \citep{cushing:2005,rayner:2009}. Then, we convolved the library spectra from G8V to M7V down to the resolution of SOFI and plotted these against each of our SOFI spectra. We estimated each SpT and a corresponding uncertainty three times by independently comparing spectra in the $J$-, $H$-, and $K$- bandpasses. The final uncertainty on each SpT corresponds to the uncertainty on the weighted mean and thus represents our best estimate of the error on this quantity. We then compute a single SpT for each star using a weighted mean. The SpT and uncertainty, rounded to the nearest tenth of a type, are listed in Table \ref{tab:stars}. Out of the 34 stars in our {\it K2} sample, we identify 27 as M dwarfs.

During our visual spectral inspection, we compared our spectra to the library spectra of giant stars in order to remove giants as early as possible in our analysis process. We identified only one star as a likely giant: EPIC 202710713, which \cite{huber:2016} and \cite{dressing:2017} also classified as an evolved star.

\subsection{Stellar Parameters}
\label{sec:star_params} 

For each absorption feature and stellar parameter (radius and effective temperature), we use least-squares fitting to determine the dependence of those parameters on the EWs calculated from the spectra. Various functional forms of EWs are used to fit the calibration sample's parameters. They include all combinations of linear, quadratic, and a ratio of EWs of two different absorption features. For example, in the simplest linear case, one lets the EW for the chosen absorption feature be the independent variable, while stellar radius or effective temperature is the dependent variable. After calculating the linear term and the offset, one then uses all the EWs to calculate the stellar radius for all the stars in our sample. This process is then repeated for all the absorption features in the spectra, all the stellar parameters, each calibration sample, and each functional combination of EWs. To account for intrinsic scatter in stellar properties, we include an additional noise term, tuned to give $\chi^{2}_{red} \approx$ 1 in the best cases. We find that scatter terms of 100 K and 0.05$R_{\odot}$ fulfill this criterion.

In order to find the optimal fit for each calibration sample, we then select the model giving the lowest Bayesian information criterion (BIC) value and the lowest scatter in the fit residuals. We use a Monte Carlo approach to estimate the uncertainties on the fit coefficients and inferred stellar parameters. Random gaussian distributions are then used to generate synthetic data sets of EWs, stellar radii, and effective temperatures. A total of 1000 trials are used for calculating the uncertainties for each parameter.

\subsection{Calibration Relations and Literature Comparison}
\label{sec:select_cal_sample}

Because some of our spectra contain residual systematics near prominent H lines, we find only poor fits using EWs located near these lines (Brackett 11-21). Viewing all possible combinations of the remaining EWs, we determine that the optimal fits for calculating our parameters are determined by having a low BIC value for the fit and comparing it to the median uncertainty of all the uncertainties in a given combination of EWs. We present the following equations for calculating stellar radius and effective temperature:
\begin{equation} 
{\frac{T_{eff}}{K} = a + b\left(\frac{Mg_{1.57}}{Al_{1.31}}\right) + c\left(\frac{Al_{1.67}}{Ca~I_{1.03}}\right)}
\end{equation}
\begin{equation} 
{\frac{R_{*}}{R_{\odot}} = a + b\left(Mg_{1.57}\right) + c\left(\frac{CO_{2.29}}{Na~I_{1.14}}\right).}
\end{equation}
Table \ref{tab:formulae} lists the best-fitting coefficients and the covariance matrix for each fit. Note that some coefficients exhibit significant correlations, suggesting that uncertainties would be underestimated if these correlations were neglected.

\begin{figure}[ht!]
\begin{center}
\includegraphics[width=3in]{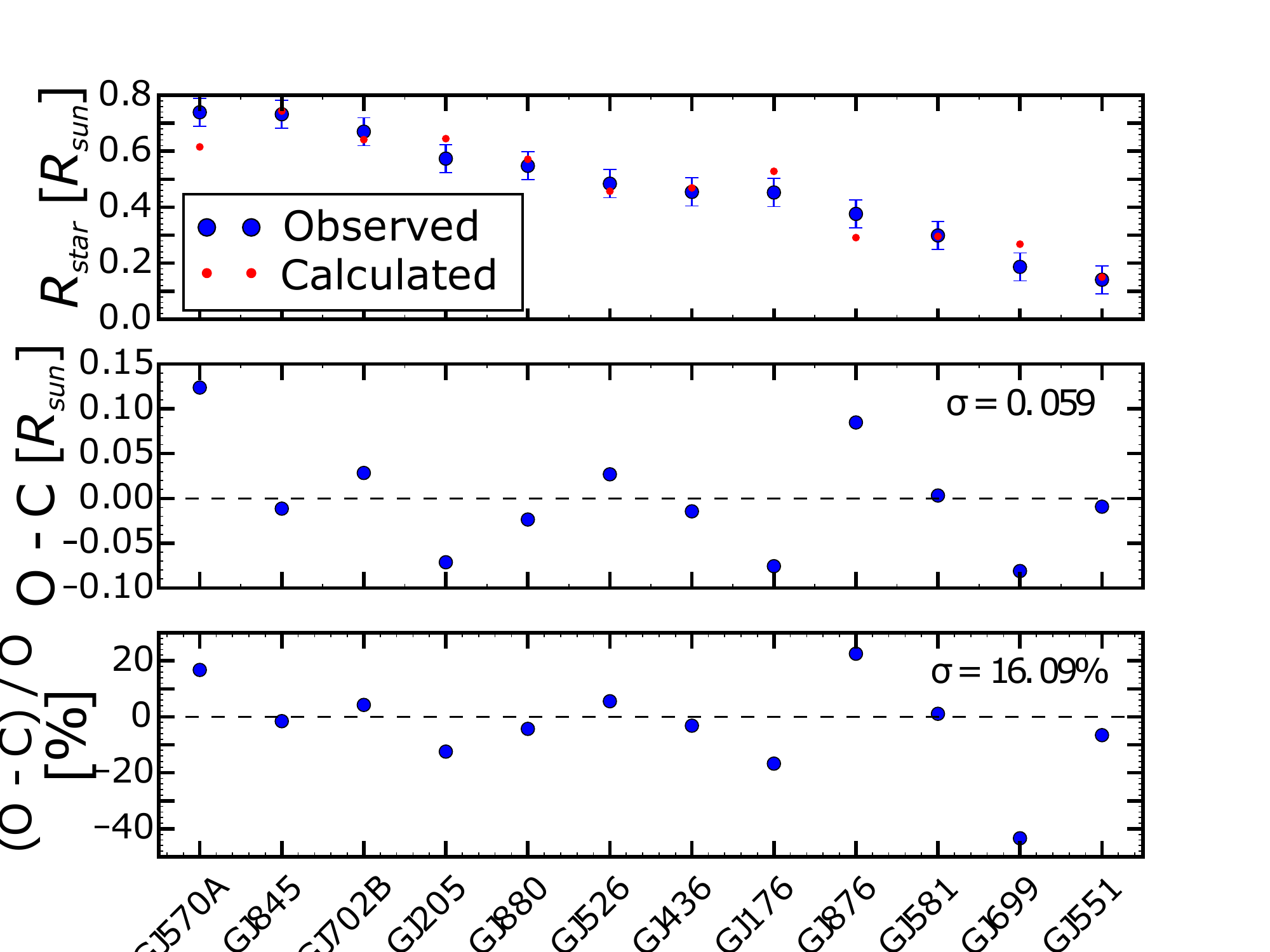}
\caption{\label{fig:radius} Stellar radius for the stars in our interferometric calibration sample, from the literature (blue circles) and derived using Eq.\ 3 (red circles).  The middle and bottom panels show the absolute and fractional deviations for each star. The dispersion of the residuals is 0.059$R_{\odot}$  and 16.09\%, respectively. Our sample spans from 0.2 to 0.7$R_{\odot}$.}
\end{center}
\end{figure}

\begin{figure}[ht!]
\begin{center}
\includegraphics[width=3in]{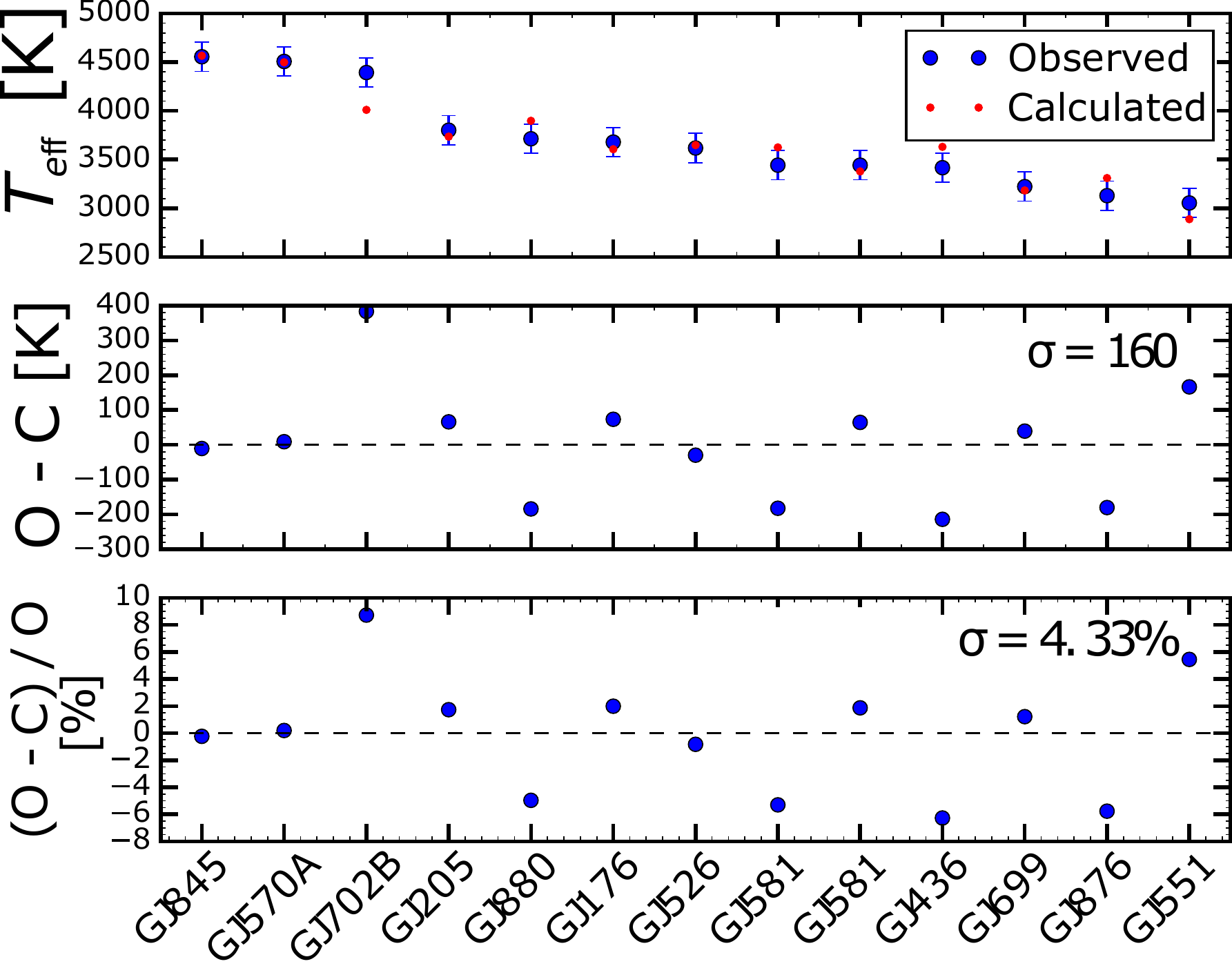}
\caption{\label{fig:teff}  Stellar effective temperature for the stars in our interferometric calibration sample, from the literature (blue circles) and derived using Eq.\ 2 (red circles).  The middle and bottom panels show the absolute and fractional deviations for each star.  The dispersion of the residuals is 160 K and 4.33\%, respectively. Our sample spans from 3000 to 4500 K.}
\end{center}
\end{figure}

\begin{figure}[ht!]
\begin{center}
\includegraphics[width=3in]{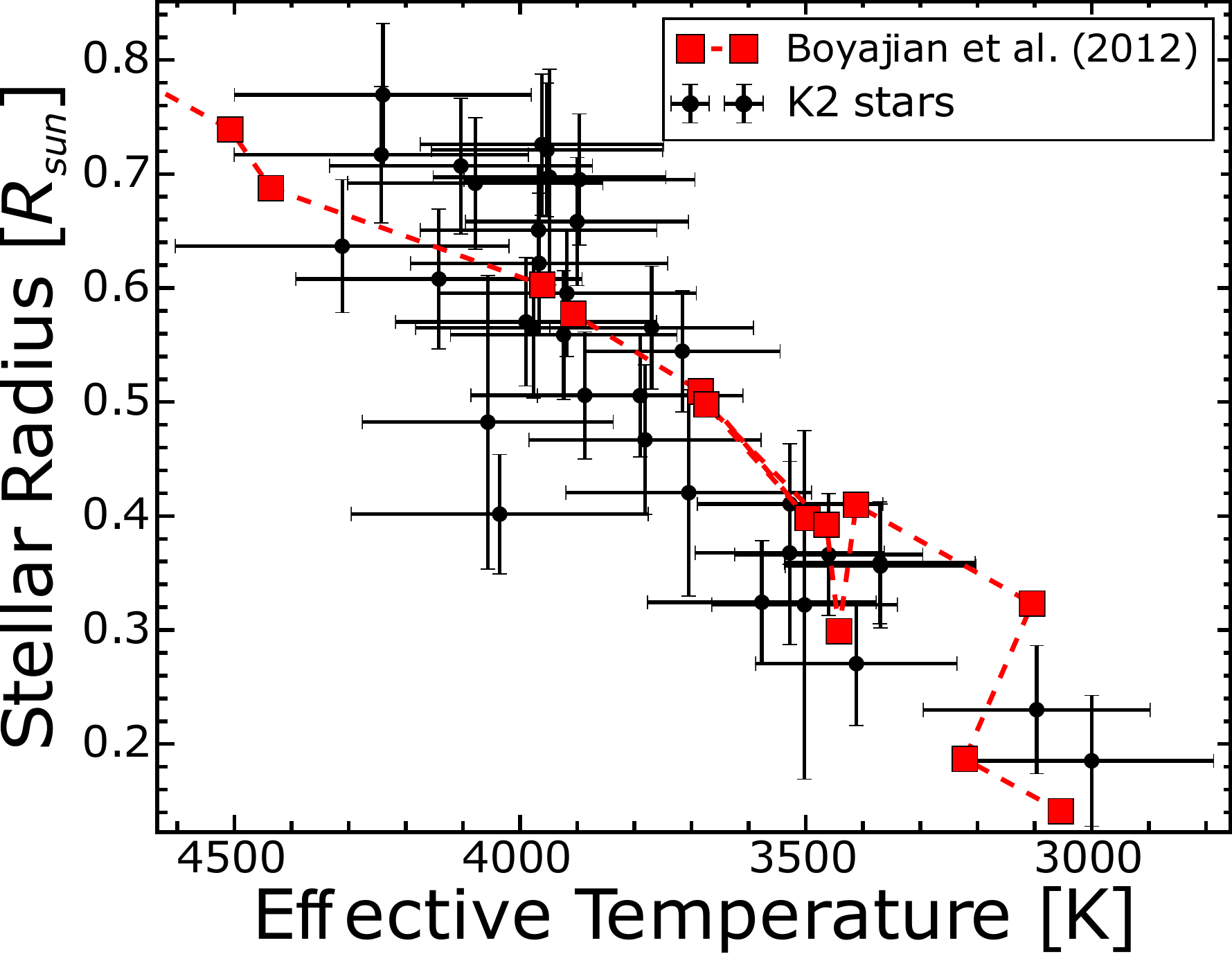}
\caption{\label{fig:stars} We show stellar radius and effective temperature for all our {\it K2} target stars (black points with error bars) derived using Eqs.\ 2 and 3. The red squares and dashed line show the average values for each SpT as calculated by \cite{boyajian:2012b}.}
\end{center}
\end{figure}

Based on the range of our calibration sample, we restrict ourselves to stars in the range 3000 K $<$ $T_{eff}$ $<$ 4500 K and 0.2 $<$ $R_{*}$/$R_{\odot}$ $<$ 0.7. There is overall excellent agreement between our derived values for radius and effective temperature, while four stars (GJ 551, GJ 699, GJ 526, GJ 876) have somewhat larger deviations in stellar radius and/or effective temperature. Figures \ref{fig:radius} and \ref{fig:teff} compare the inferred and literature values for our calibrated sample. The middle and bottom panels of these two figures show that the dispersions of the residuals are 0.059$R_{\odot}$ (16.09\%) and 160 K (4.33\%) for stellar radius and effective temperature, respectively. All of the stars in our calibration sample, with the exception of GJ 526, have published luminosities (calculated using the Stefan-Boltzmann law) within 1$\sigma$ of our inferred values. Finally, we estimate each star's mass by inverting the mass-radius relationship of \cite{maldonado:2015}. The full set of stellar values is listed in Table \ref{tab:stars} and the {\it K2} stellar parameters are plotted in Figure \ref{fig:stars}.

We also independently compare our stellar parameters to those of \cite{dressing:2017}. Out of our 34-star {\it K2} sample (as referenced in Table \ref{tab:stars}), we share 21 stars in common with their sample. While this work calculates stellar parameters using the spectra acquired with NTT/SOFI, \cite{dressing:2017} use two different instruments in their work. The SpeX instrument, on the NASA Infrared Telescope Facility, provides wavelength coverage from 0.7 to 2.55 $\micron$ at a resolution of R $\approx$ 2000 \citep{rayner:2003}. The other instrument used was TripleSpec on the Palomar 200", providing wavelength coverage from 1.0 to 2.4 $\micron$ at a resolution of R $\approx$ 2500-2700 \citep{herter:2008}. \cite{dressing:2017} derive and compare stellar parameters using EW-based relations developed by \cite{newton:2015} and index-based relations from \cite{mann:2013b}. Both sets of relations were calibrated using a set of stars with interferometrically determined parameters from \cite{boyajian:2012b}. Ultimately, effective temperatures, stellar radii, and luminosities were derived using the \cite{newton:2015} relations, stellar masses\footnote{Using the effective temperatures from the \cite{newton:2015} relation} and metallicity were calculated using the \cite{mann:2013b} relations, and surface gravities were calculated from masses and stellar radii. 

\begin{figure}[ht!]
\begin{center}
\includegraphics[width=3in]{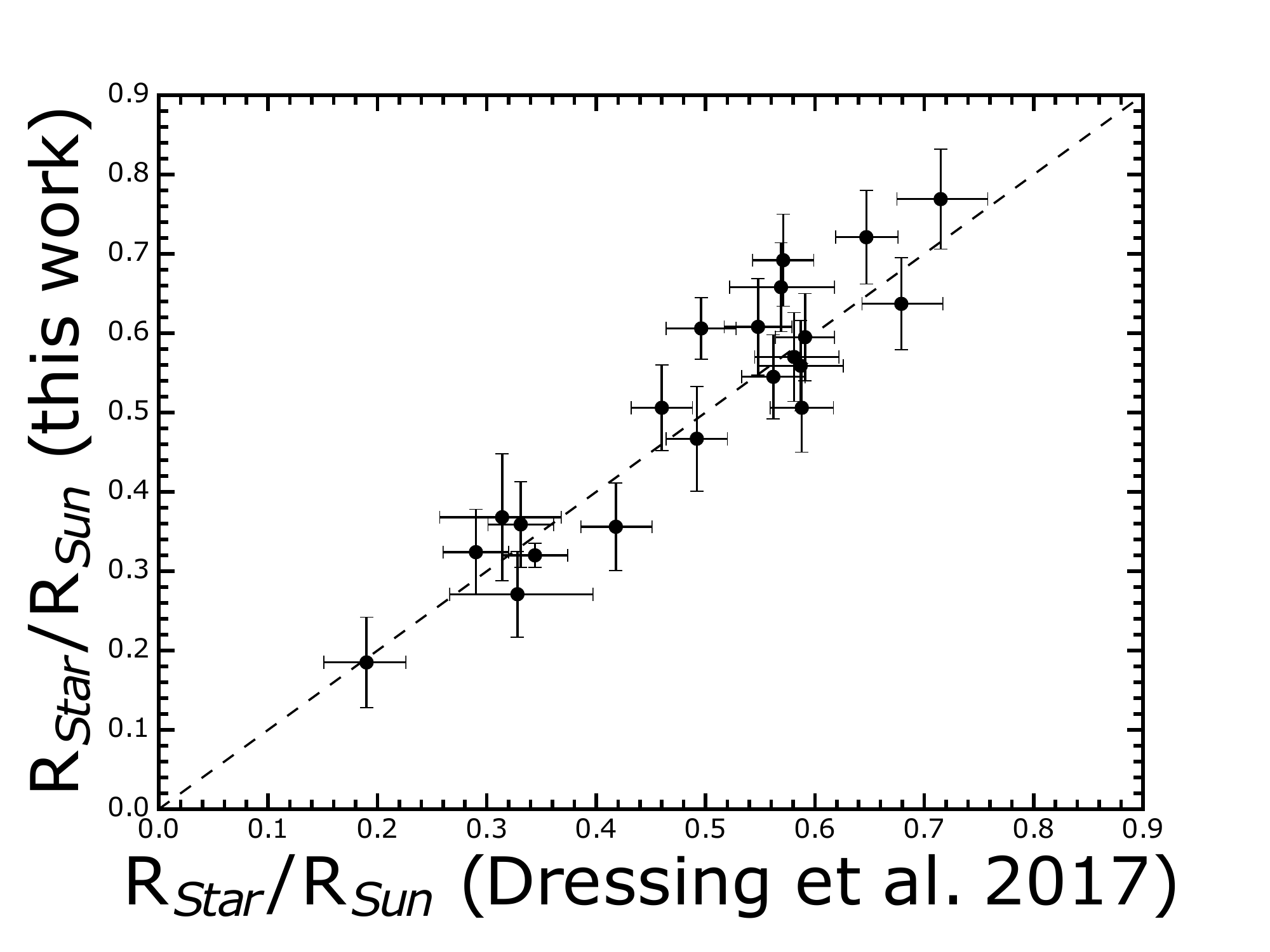}
\caption{\label{fig:mtz_dressing_rstar} Comparison of our stellar radii to those of \cite{dressing:2017}. The dotted line shows a 1:1 agreement, while any deviation from the dotted line presents the small discrepancies. Overall, there is a general agreement between our works in deriving our stellar radii.}
\end{center}
\end{figure}

\begin{figure}[ht!]
\begin{center}
\includegraphics[width=3in]{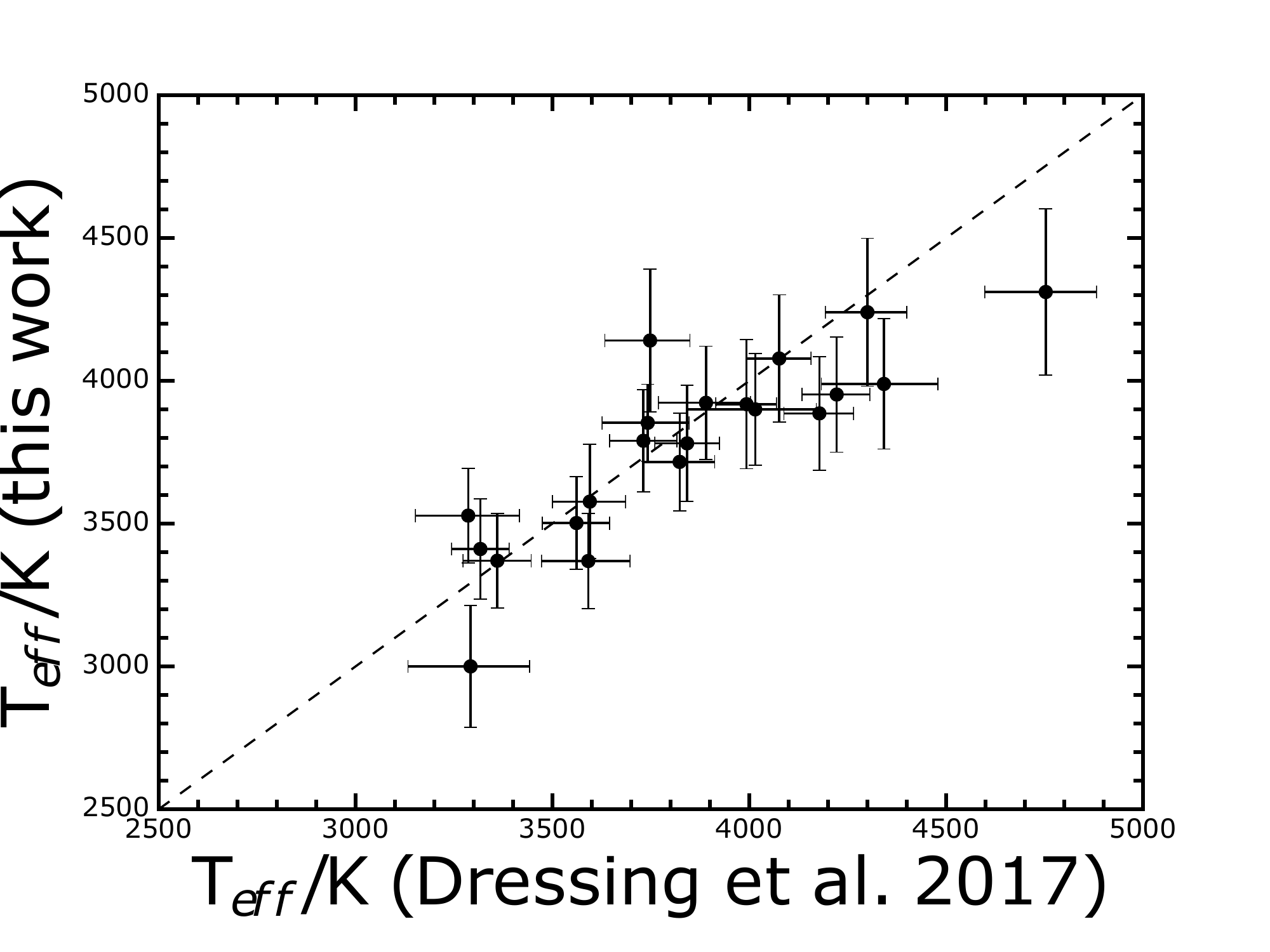}
\caption{\label{fig:mtz_dressing_teff} Comparison of our effective temperatures to those of \cite{dressing:2017}. The dotted line shows a 1:1 agreement, while any deviation from the dotted line presents the small discrepancies. Overall, there is a general agreement between our works in deriving our effective temperatures. We address small caveats in \S\ref{sec:select_cal_sample} for the outlier, EPIC 211770795.}
\end{center}
\end{figure}

Comparing the parameters derived by \cite{dressing:2017} with those shown in Figures \ref{fig:mtz_dressing_rstar} and \ref{fig:mtz_dressing_teff}, we find $\chi^{2}_{red}$ < 1 in both cases. This indicates that there is an excellent agreement between our two methods and verifies the validity the of our approach. Additionally, our stellar parameters are consistent with those from a number of previous publications \citep{crossfield:2015a, montet:2015, petigura:2015a, mann:2016, obermeier:2016, schlieder:2016}.

The most highly discrepant system evident in Figure \ref{fig:mtz_dressing_teff} seems to be the effective temperature of EPIC 211770795. Our estimate is significantly lower than the 4750 K estimated by \cite{dressing:2017}. Their value is larger than the 4500 K upper limit determined from our calibration sample (see Figure \ref{fig:radius}), providing further evidence that our relations are not well calibrated beyond this range. Furthermore, we see an offset between our effective temperature values and those reported by \cite{dressing:2017}, demonstrating that systematic calibration errors may still play a role in one or both of these analyses.  As seen with the index-based relations of \cite{mann:2015}, our EW-based relations also start to saturate around 4000 K and could systematically effect any derived planetary parameters, such as equilibrium temperatures.

Metallicity could be a factor for some stars and could cause a shift in effective temperature and stellar radius. The larger uncertainties in our stellar parameters when compared to those of \cite{dressing:2017} may result from a range of stellar metallicities.

\begin{figure}[ht!]
\begin{center}
\includegraphics[width=3.25in]{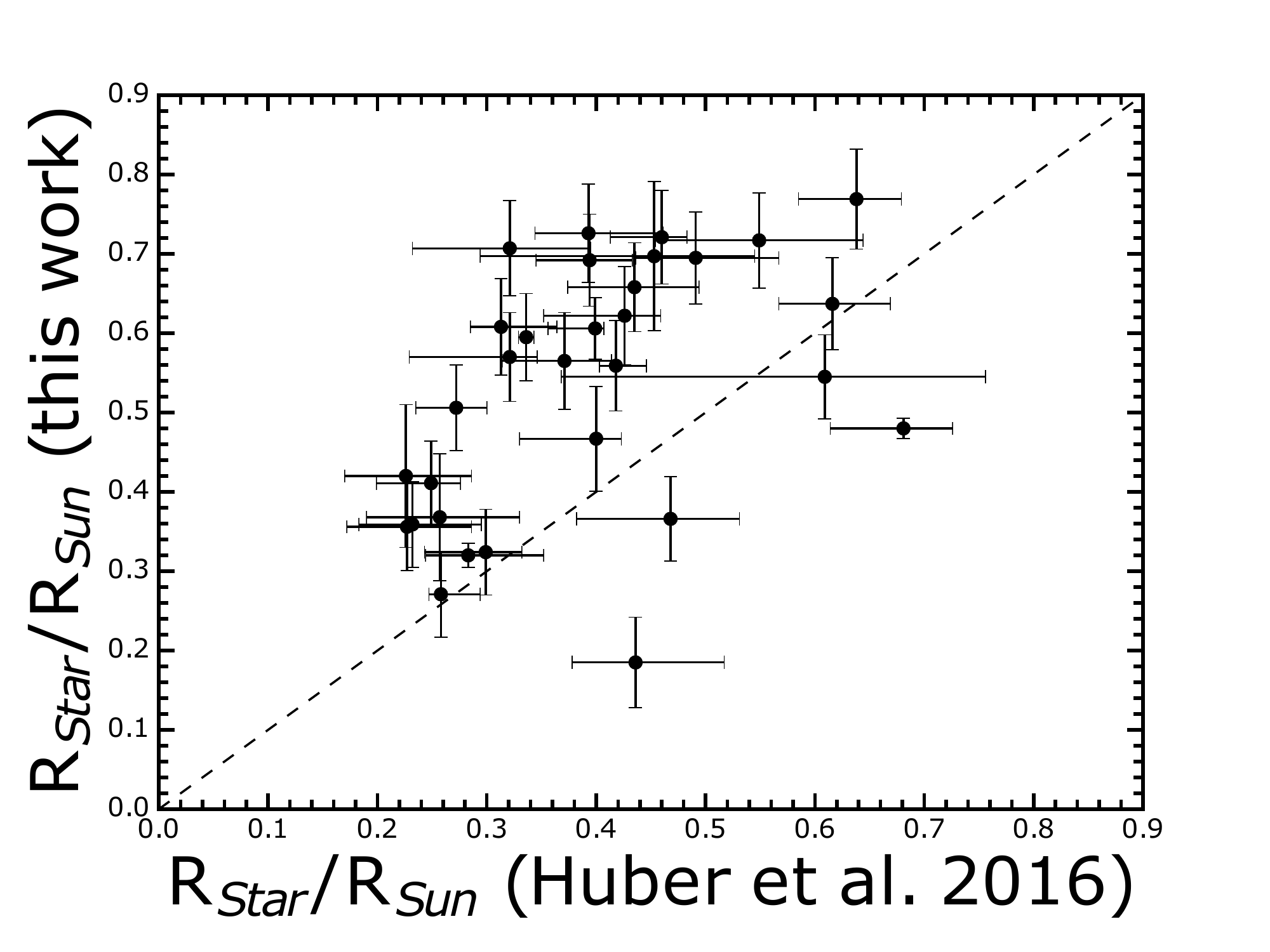}
\caption{\label{fig:hub_mtz_rstar} Comparison of our stellar radii to those of \cite{huber:2016}. The dotted line shows a 1:1 agreement, while any deviation from the dotted line presents the small discrepancies. As in discussed in \S\ref{sec:select_cal_sample}, we find that the majority of the objects in the sample are larger in our work and find a 0.15 $R_{\odot}$ median increase.}
\end{center}
\end{figure}

\begin{figure}[ht!]
\begin{center}
\includegraphics[width=3.25in]{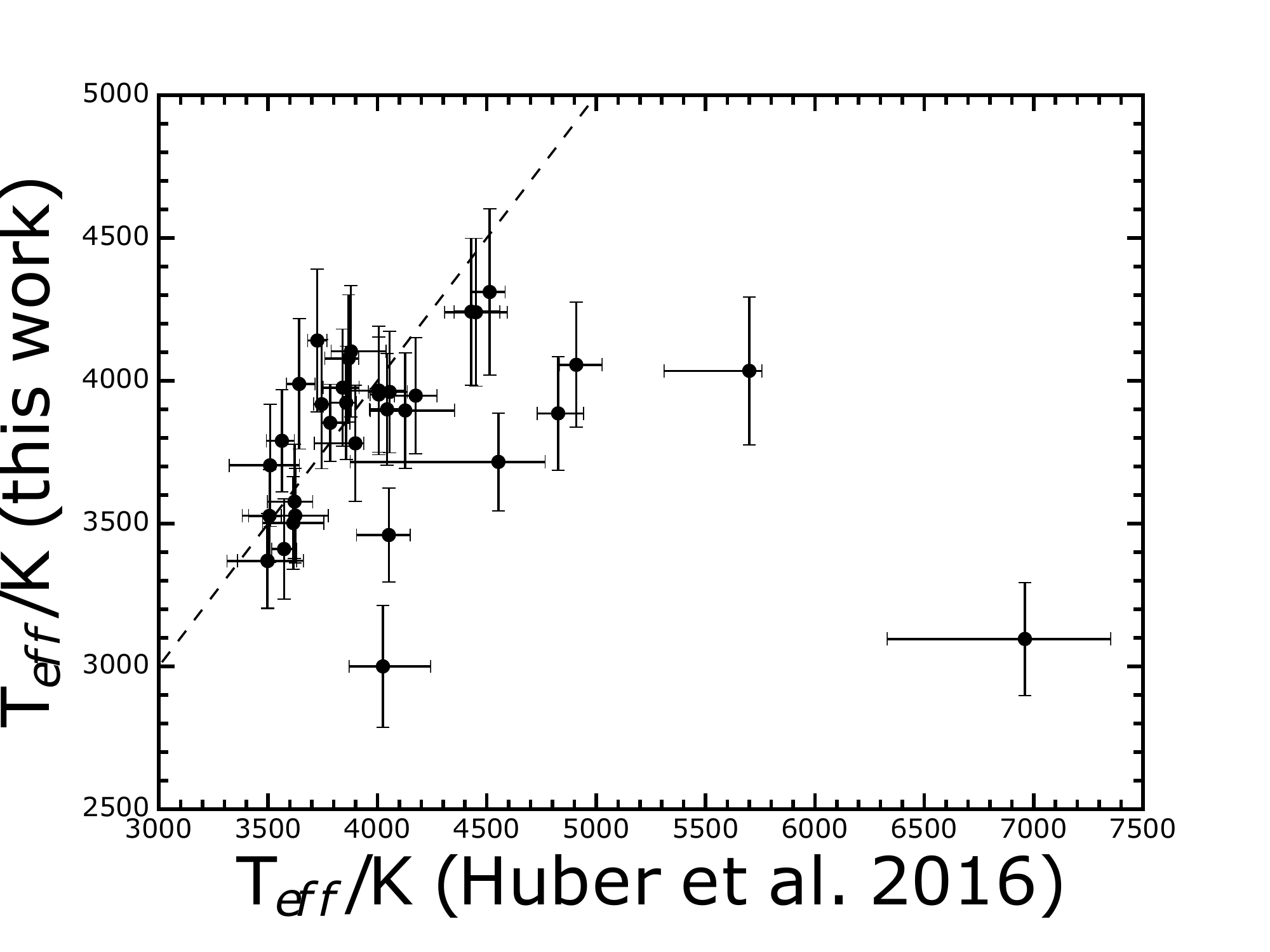}
\caption{\label{fig:hub_mtz_teff} Comparison of our in effective temperatures to those of \cite{huber:2016}. The dotted line shows a 1:1 agreement, while any deviation from the dotted line presents the small discrepancies. See \S\ref{sec:select_cal_sample} for a discussion.}
\end{center}
\end{figure}

Additionally, we compare all 34 of our stellar parameters with the photometrically derived stellar parameters from \cite{huber:2016}, shown in Figures \ref{fig:hub_mtz_rstar} and \ref{fig:hub_mtz_teff}. Figure \ref{fig:hub_mtz_rstar} shows that there is a median increase of 0.15$R_{\odot}$ when comparing our stellar radii to those of \cite{huber:2016}. Figure \ref{fig:hub_mtz_teff} shows a general agreement in effective temperature between both of our works with the exception of EPIC 204489514 and EPIC 205145448.

We note that the analysis done in \cite{huber:2016} is subject to the limitations of broadband photometry. Furthermore, \cite{huber:2016} note that model-based estimates tend to underpredict stellar radii by 20\% \citep{boyajian:2012a} and encourage the use of empirical calibrations for estimating the stellar parameters in cool dwarfs. Lastly, our empirically calculated parameters are in agreement with those in \cite{dressing:2017} for the points where we disagree with the values of \cite{huber:2016}, giving us further confidence in our results.

\subsection{Planetary Parameters}
\label{sec:planet_params}
Radii and equilibrium temperatures of transiting planets are calculated using the stellar parameters of its host star. Using the transit depths and periods measured using {\it K2} photometry~\citep{crossfield:2016} and our newly calculated stellar parameters, planet radii are determined with the following equation:
\begin{equation} 
{\Delta L = \left(\frac{R_{p}}{R_{*}}\right)^{2}}
\end{equation}
where $\Delta L$ is the transit depth of the planet candidate with respect to its host star.

\begin{figure}[ht!]
\begin{center}
\includegraphics[width=3in]{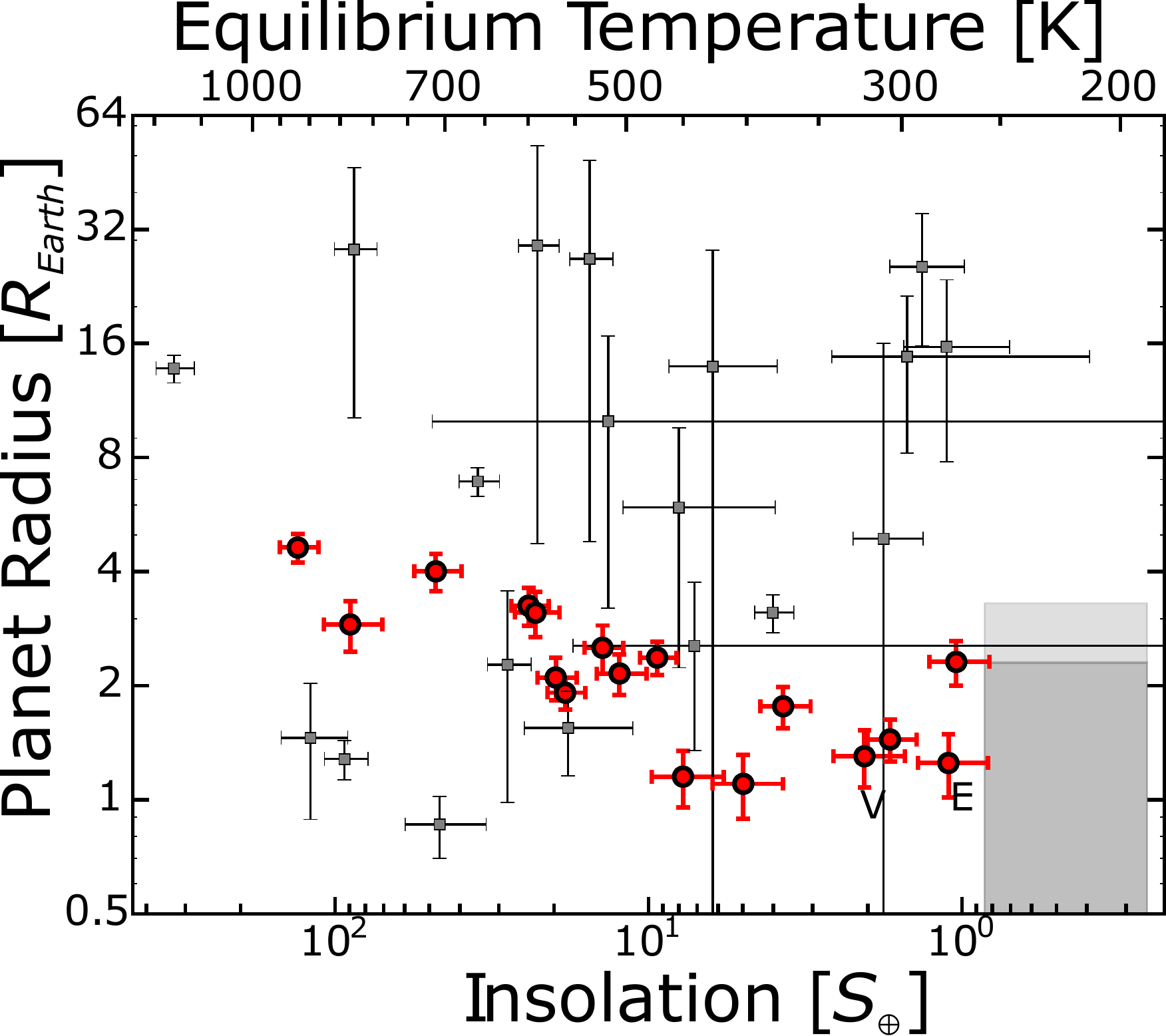}
\caption{\label{fig:planets} Planet radii, incident irradiation, and equilibrium temperatures of all {\it K2} planets and candidates observed in our program.  Venus and Earth are indicated by single letters. Plus red signs indicate validated planets, and gray squares indicate planet candidates, as reported by~\cite{crossfield:2016}. The shaded region represents the approximate location of the cloud-free habitable zone for an early-type M dwarf \citep{kopparapu:2013a}. That zone was defined for planets with masses 0.3--10 times that of Earth. The larger of those masses corresponds to the upper, lightly shaded area \citep{wolfgang:2016}.}
\end{center}
\end{figure}

Calculating the equilibrium temperature of a planet candidate requires more parameters from the planet and its host star. The following equation calculates the equilibrium temperatures for each {\it K2} planet or candidate as a comparison to our own Earth-Sun system:
\begin{equation} 
{T_{eq} = (270 K) \left( \frac{T_{eff}}{T_{eff,\odot}}\right) \left(\left[ \frac{R_{*}}{R_{\odot}} \right]\left[\frac{1 AU}{a}\right]\right)^{1/2} }
\end{equation}
where $T_{eff}$ is the effective temperature of the star, $R_{*}$ is the radius of the star, and {\it a} is the semi-major axis of the planet orbiting its parent star. Here we calculate the semi-major axis of the planet by using Kepler's third law. The 270 K equilibrium temperature scaling factor corresponds to a Bond albedo of 0.3, which is comparable to that inferred for gas giants more highly irritated than Earth. All uncertainties are propagated through the entire calculation for planet radii. We present the derived values for our {\it K2} planets and planet candidates in Table \ref{tab:candidates} and plot these derived values (along with incident irradiation) in Figure \ref{fig:planets}.

Our sample shown in Figure \ref{fig:planets} includes 18 validated planets and 19 remaining planet candidates. While fitting for the light-curve parameters of these remaining candidates, degeneracies (such as impact parameters near unity) arose that preclude any precise determination of $R_{p}$/$R_{*}$. The candidates have much larger uncertainties on their size, which typically makes statistical validation much more difficult. Based on the paucity of large (>6$R_{\oplus}$) planets orbiting M dwarfs \citep{johnson:2007, johnson:2010}, the $\gtrsim$ 9 candidates larger than this size are likely false positives; since planet validation is not the aim of this work, we retain the previously assigned designation of planet candidate. 

In addition to these likely false positives, our validated planets include several hot Neptunes and two planets (K2-18b and K2-72e) that lie near the habitable zone. Of our whole {\it K2} sample, only eight planets (three of which are still planet candidates) are smaller than 1.6 Earth radii. According to \cite{rogers:2015}, planets smaller than 1.6 Earth radii are likely to have compositions dominated by rock or iron, while larger planets are more likely to be volatile-rich. However, there may still be rocky planets larger than this limit. For example, \cite{buchhave:2016} found that {\it Kepler}-20b, a 1.9$R_{\oplus}$ planet, has a density consistent with a rocky composition even though it is beyond the rocky-to-gaseous transition.

\begin{figure}[ht!]
\begin{center}
\includegraphics[width=3.5in]{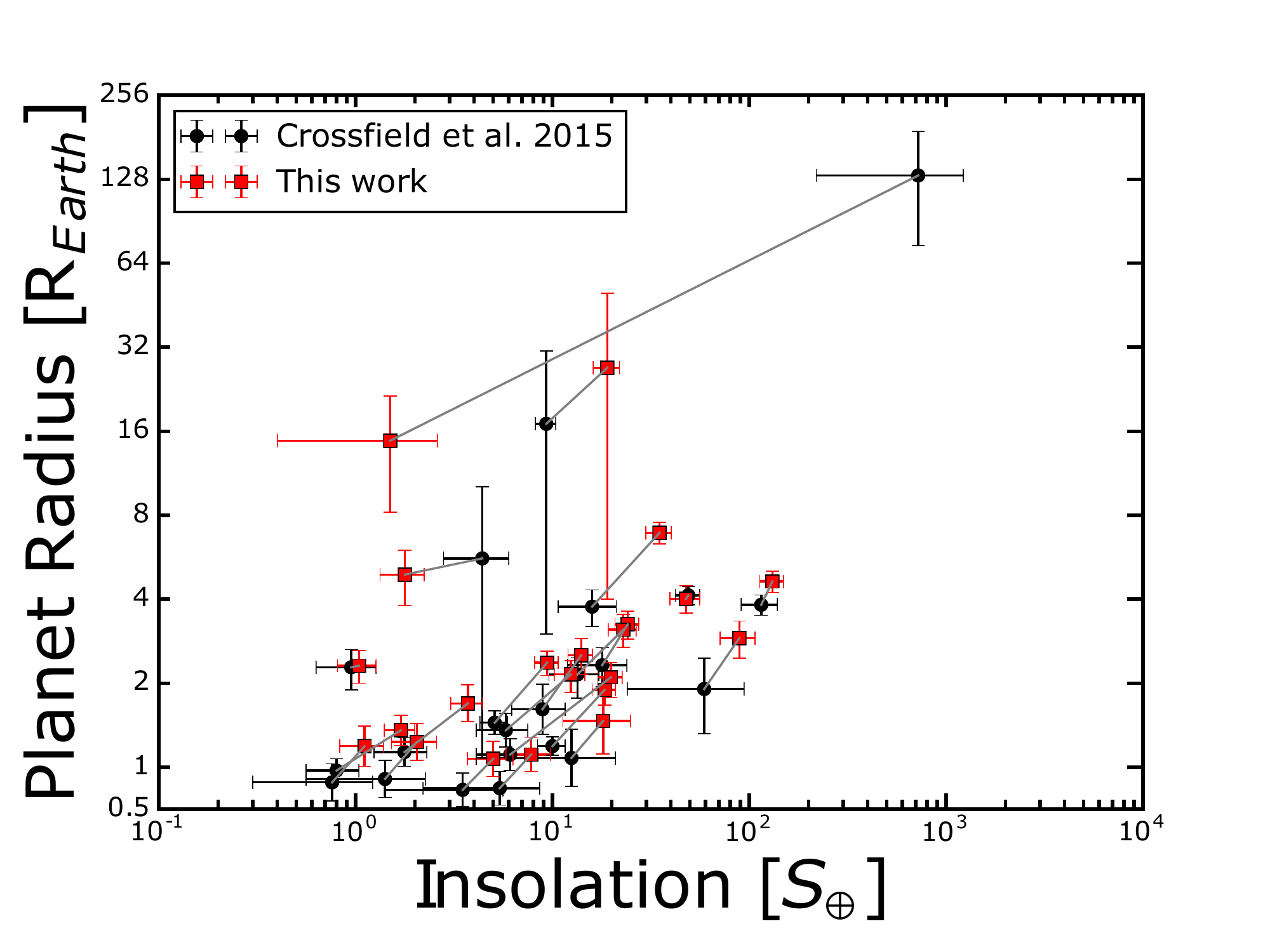}
\caption{\label{fig:mtz_crossfield_teq} Planet radii and incident irradiation for the {\it K2} planets and planet candidates that appear in both our work and \cite{crossfield:2016}. Black circles indicate the {\it K2} objects reported by \cite{crossfield:2016}, while the dark red squares indicate the {\it K2} objects in this work. The dark gray lines connect our updated parameter values to the original estimates published by \cite{crossfield:2016}. See \S\ref{sec:planet_params} for details.}
\end{center}
\end{figure}

We compare our calculations of the insolation flux from our {\it K2} sample to those from \cite{crossfield:2016} in Figure \ref{fig:mtz_crossfield_teq}. The discrepancies between our values and those in \cite{crossfield:2016} highlight the importance of using spectroscopically derived stellar parameters in order to compute planet parameters.

\section{Conclusion and Future Prospects}
\label{sec:conc_future}
In this paper, we derive stellar and planetary parameters for {\it K2} K and M dwarf systems. We adopt similar calibration techniques from \cite{neves:2014}, \cite{maldonado:2015}, and \cite{terrien:2015} by measuring EWs in the near-infrared part of the spectrum. Interferometric calibration samples are used from~\cite{demory:2009},~\cite{vonbraun:2011,vonbraun:2012},~\cite{boyajian:2012b}, and~\cite{vonbraun:2014} in order to provide a more precise baseline to calculate the stellar radii and effective temperatures of the stars in our sample. Various functions (whether they are linear, quadratic, or a ratio of EWs) are tested, and we use the functions with the best BIC value and the lowest residuals to calculate stellar parameters. 

Our spectroscopically derived stellar radii improve on previously reported values that relied on stellar models poorly calibrated to these low-mass stars. We find a median increase of 0.15$R_{\odot}$ when comparing our measurements to those of \cite{huber:2016}, consistent with the median increase in size found by \cite{newton:2015} when revising the photometrically based stellar radius estimates determined by \cite{dressing:2013} for cool dwarfs observed during the prime Kepler mission. Finally, we calculate the {\it K2} planet or planet candidate radius and equilibrium temperature. 

Since our team also obtained optical spectra, using the EFOSC2 spectrograph \citep{buzzoni:1984} on the NTT, in a future work we will apply the same techniques in order to cross-check our stellar properties. Furthermore, this work does not calculate stellar metallicities; however, we plan to so in later works.

Our work paves the way for future exoplanet surveys. Other spectroscopic and photometric surveys focusing on M dwarfs are currently underway or are being planned for the near future. SPECULOOS, a 1 m near-infrared telescope, will observe approximately 500 of the nearest M and brown dwarfs in the southern hemisphere \citep{gillon:2013b}. CARMENES will provide high-resolution (R = 82,000) spectra between 0.5 and 1.7 \micron~for late-type M dwarfs and search for Earth-like planets in the habitable zone \citep{quirrenbach:2012}. The Habitable Zone Planet Finder (HZPF) will also provide spectra for M dwarfs and will attempt to find planets through the Doppler effect \citep{mahadevan:2010}. Yet another RV survey, SPIRou, aims to find exoplanets around low-mass stars using high-resolution spectra between 0.98 and 2.35 \micron~\citep{santerne:2013}.

Future transit surveys will detect many new Earth-like planets around M dwarfs, just like previous and ongoing photometric surveys such as {\it Kepler} and {\it K2}. Although the current Gaia mission \citep{lindegren:2010} focuses more on astrometry (for which stellar mass is a key input), its two photometers can provide light curves for exoplanet detection. The {\it Transiting Exoplanet Survey Satellite} \citep[TESS;][]{ricker:2009} and {\it PLAnetary Transits and Oscillations of stars} \citep[PLATO;][]{rauer:2014} will also find planets, some of which will be high-priority targets for the {\it James Webb Space Telescope} \citep[JWST;][]{gardner:2006}. The recent announcement of a roughly Earth-mass planet candidate orbiting Proxima Centauri \citep{anglada-escude:2016} adds yet more urgency to the need to search for more planets and characterize their low-mass host stars. The combination of all of these surveys will yield many new M dwarf systems in need of stellar and planetary parameters and of a large, precise calibration sample.

\acknowledgments{A.~O.~M.~would like to thank all of the members of the {\it K2} team for all the assistance and interesting conversations throughout this work. A.~W.~H.\ acknowledges support for our {\it K2} team through a NASA Astrophysics Data Analysis Program grant.  A.~W.~H.\ and I.~J.~M.~C.\ acknowledge support from the {\it K2} Guest Observer Program. Finally, we thank the anonymous referee for the insightful comments that improved the quality of this manuscript.

This material is based on work supported by the National Science Foundation under Award nos. AST-1322432, a PAARE Grant for the California-Arizona Minority Partnership for Astronomy Research and Education (CAMPARE), and DUE-1356133, an S-STEM Grant for the Cal-Bridge CSU-UC PhD Bridge Program. This work was funded in part by Spitzer GO 11026 (PI Werner), managed by JPL/Caltech under a contract with NASA and locally by the University of Arizona. This work was performed in part under contract with the California Institute of Technology/Jet Propulsion Laboratory funded by NASA through the Sagan Fellowship Program executed by the NASA Exoplanet Science Institute. Travel costs were partially supported by the National Geographic Society. Any opinions, findings, and conclusions or recommendations expressed in this material are those of the author(s) and do not necessarily reflect the views of the National Science Foundation.}

\bibliographystyle{apj}

\clearpage
\clearpage
\begin{deluxetable}{llllllll}
\tabletypesize{\scriptsize}\tablecaption{Stellar Calibration Sample}\tablehead{
      Star &  SpT \tablenotemark{a}  &  $R_*$ &  $T_{eff}$ &   $L_*$ & Reference & Notes \\
      	& & [$R_{\odot}$]  & [K] & [$L_{\odot}$] & \\
}\startdata
     GJ176 & M2.5V & 0.453(22)  &  3679(77) & 0.0337(43)    &    \cite{vonbraun:2014}  &    \\
     GJ205 &  M1.5V & 0.5735(44) &  3801(9)  & 0.0616(11)    &    \cite{boyajian:2012b}   &   \\
     GJ436 & M3V & 0.455(18)  &  3416(53) & 0.0253(25)    &    \cite{vonbraun:2012}  &  1  \\
     GJ526 & M1.5V & 0.4840(84) &  3618(31) & 0.0360(18)    &    \cite{boyajian:2012b}   &   \\
     GJ551 & M5.5V & 0.1410(70) &  3054(79) & 0.00155(22)   &    \cite{boyajian:2012b}   &   \\
    GJ570A & K4V & 0.739(19)  &  4507(58) & 0.202(15)     &    \cite{demory:2009}  &    \\
     GJ581 & M2.5V & 0.299(10)  &  3442(54) & 0.0113(10)    &    \cite{vonbraun:2011}  &  2  \\
     GJ699 & M4.0V & 0.1869(12) &  3222(10) & 0.003380(60)  &    \cite{boyajian:2012b}  &   \\
    GJ702B & K5Ve & 0.6697(89) &  4400(150)& 0.150(46)     &    \cite{boyajian:2012b}  &    \\
     GJ845 & K5V & 0.7320(60) &  4555(24) & 0.207(34)     &    \cite{demory:2009}  &    \\
     GJ876 & M3.5V & 0.3761(59) &  3129(19) & 0.0122(39)    &    \cite{vonbraun:2014}  &  3  \\
     GJ880 & M1.5V & 0.5477(48) &  3713(11) & 0.0512(90)    &    \cite{boyajian:2012b}  &    \\
\enddata\label{tab:calibration}
\tablenotetext{a}{Spectral types were adopted from the interferometric works, with the 
following exceptions: (1) \cite{kirkpatrick:1991,hawley:1996}; (2) \cite{henry:1994}; and (3)
 were linearly interpolated from \cite{pickles:1998}.}
\clearpage
\end{deluxetable}

\clearpage
\begin{deluxetable}{l r r r r r r}
\tabletypesize{\scriptsize}\tablecaption{\label{Tab:lines}\ $J$-, $H$-, and $K$-band equivalent width features}
\tablecolumns{7}
\tablehead{ \colhead{Feature} & \multicolumn{2}{c}{Feature window} & \multicolumn{2}{c}{Blue continuum} & \multicolumn{2}{c}{Red continuum} \\
& \multicolumn{2}{c}{\micron} & \multicolumn{2}{c}{\micron} & \multicolumn{2}{c}{\micron}}
\startdata
Ca I ($1.03\micron$) & 1.0320 & 1.0365 & 1.0280 & 1.0315 & 1.0368 & 1.0377 \\
Na I ($1.14\micron$) & 1.1361 & 1.1432 & 1.1270 & 1.1327 & 1.1478 & 1.1572 \\
Al ($1.31\micron$) & 1.3125 & 1.3180 & 1.3060 & 1.3090 & 1.3180 & 1.3220 \\
Mg ($1.48\micron$) & 1.4865 & 1.4905 & 1.4810 & 1.4850 & 1.4920 & 1.4960  \\
Mg ($1.50\micron$) & 1.5002 & 1.5075 & 1.4910 & 1.4983 & 1.5090 & 1.5163  \\
Mg ($1.57\micron$) & 1.5725 & 1.5797 & 1.5665 & 1.5720 & 1.5810 & 1.5865  \\
Si ($1.58\micron$) & 1.5875 & 1.5925 & 1.5820 & 1.5865 & 1.5930 & 1.5975 \\
CO ($1.62\micron$)  & 1.6178 & 1.6280 & 1.6048 & 1.6150 & 1.6300 & 1.6402  \\
Al ($1.67\micron$)  & 1.6698 & 1.6790 & 1.6558 & 1.6650 & 1.6800 & 1.6892 \\
Mg ($1.71\micron$) & 1.7089 & 1.7139 & 1.7000 & 1.7050 & 1.7149 & 1.7199  \\
Na I ($2.20\micron$) & 2.2020 & 2.2120 & 2.1890 & 2.1990 & 2.2125 & 2.2225 \\
Ca I ($2.26\micron$) & 2.2586 & 2.2696 & 2.2480 & 2.2570 & 2.2700 & 2.2800 \\
CO ($2.29\micron$) & 2.292 & 2.315 & 2.286 & 2.290 & 2.315 & 2.320 \\
\enddata
\tablecomments{All the wavelengths are presented at their rest wavelength.}
\clearpage
\end{deluxetable}

\begin{deluxetable}{llllll}
\tabletypesize{\scriptsize}\tablecaption{Equivalent Width Formulae}\tablehead{
      Quantity &  Formula & a & b & c \\
}\startdata
$T_{eff}$ & $a + b(Mg_{1.57}/Al_{1.31}) + c(Al_{1.67}/Ca~I_{1.03})$      & 2989.5 & -577.05 & 53.804\\
 & uncertainties: & 78.56147 & 52.42034 & 10.44419 \\
 & Covariance: & & & \\
 & & 6171.9 & -3355.2 & -493.87 \\
 & & -3355.2 & 2747.9 & 265.68 \\
& & -493.87 & 265.68 & 109.08 \\
\\
$R_*$    & $a + b(Mg_{1.57}\AA) + c(CO_{2.29}/Na~I_{1.14})$ & 0.18552 & 1265.2 & 0.010852\\
 & uncertainties: & 0.02569482 & 117.2119 & 0.005063553 \\
 & Covariance: & & & \\
 & & 0.00066022 & -1.8673 & -0.00003695 \\
& & -1.8673 & 13739 & -0.2305 \\
& & -0.00003695 & -0.2305 & 0.00002564 \\
\enddata\label{tab:formulae}
\end{deluxetable}

\clearpage
\begin{turnpage}
\begin{deluxetable}{lllllllllllll}


\tablewidth{1.22\textwidth}
\tabletypesize{\scriptsize}\tablecaption{Derived Stellar Parameters}\tablehead{
        Star \tablenotemark{a} &         SpT &      $R_*$ &  $T_{eff}$ &      $M_*$ &        $L_*$ & $\alpha_{2000}$ & $\delta_{2000}$ & $Kp$ & $J$& $H$ & $K$ &Notes \tablenotemark{b} \\
               &     &  [$R_\odot$] &  [K]  &    [$M_\odot$] &    [$L_\odot$] & & & [mag] & [mag] & [mag] & [mag] & \\
}\startdata
       GJ176 &   M2.4(0.7) &  0.528(53) &  3605(170) &  0.546(55) &   0.0423(63) & $04^{h}42^{m}55\fs774$ & $+18\arcdeg57\arcmin29\farcs404$ & -- & 6.462 & 5.824 & 5.607 & 1 \\
       GJ205 &   M1.4(1.1) &  0.645(56) &  3735(172) &  0.665(55) &   0.0726(96) & $05^{h}31^{m}27\fs395$ & $-03\arcdeg40\arcmin38\farcs031$ & -- & 4.83  & 4.05  & 3.90  & 2 \\
       GJ436 &   M2.5(0.6) &  0.469(53) &  3630(165) &  0.484(57) &   0.0343(57) & $11^{h}42^{m}11\fs094$ & $+26\arcdeg42\arcmin23\farcs65 $ & -- & 6.900 &  6.319 & 6.073 & 3 \\
       GJ526 &   M2.3(1.0) &  0.457(53) &  3647(173) &  0.472(57) &   0.0332(56) & $13^{h}45^{m}43\fs776$ & $+14\arcdeg53\arcmin29\farcs463$ & -- & 5.18  & 4.78  & 4.415 & 2 \\
       GJ551 &   M4.8(0.6) &  0.150(59) &  2887(234) &  0.104(78) &  0.00141(79) & $14^{h}29^{m}42\fs948$ & $-62\arcdeg40\arcmin46\farcs163$ & -- & 5.357 & 4.835 & 4.384 & 4 \\
       GJ570A &   K4.1(0.7) &  0.615(55) &  4498(417) &  0.636(55) &   0.139(22) & $14^{h}57^{m}28\fs001$ & $-21\arcdeg24\arcmin55\farcs713$ & -- & 3.83  & 3.23  &  3.10  & 4 \\
       GJ581 &   M2.5(0.7) &  0.279(60) &  3624(208) &  0.266(72) &   0.0121(37) & $15^{h}19^{m}26\fs823$ & $-07\arcdeg43\arcmin20\farcs21 $ & -- &  6.706 &  6.095 & 5.837 & 5 \\
       GJ699 &   M3.1(0.8) &  0.268(55) &  3182(185) &  0.253(68) &   0.0066(20) & $17^{h}57^{m}48\fs498$ & $+04\arcdeg41\arcmin36\farcs207$ & -- & 5.244 & 4.83  & 4.524 & 2, 6 \\
       GJ702B &   M3.5(0.6) &  0.641(56) &  4009(225) &  0.661(56) &    0.095(13) & $18^{h}05^{m}27\fs421$ & $+02\arcdeg29\arcmin56\farcs42$ & -- & -- & -- & -- & 2 \\
       GJ845 &   M1.1(2.3) &   0.74(16) &  4565(706) &   0.76(15) &    0.216(72) & $22^{h}03^{m}21\fs658$ & $-56\arcdeg47\arcmin09\farcs516$ & -- & 2.894 & 2.349 & 2.237 & 4 \\
       GJ876 &   M2.4(0.6) &  0.291(54) &  3309(170) &  0.282(65) &   0.0091(24) & $22^{h}53^{m}16\fs733$ & $-14\arcdeg15\arcmin49\farcs318$ & -- & 5.934 &  5.349 & 5.010 & 1 \\
       GJ880 &   M0.7(1.0) &  0.571(54) &  3897(193) &  0.591(55) &   0.0676(97) & $22^{h}56^{m}34\fs804$ & $+16\arcdeg33\arcmin12\farcs354$ & -- & 5.360 &  4.800 &  4.523 & 2 \\
   201205469 &   M0.9(1.0) &  0.559(57) &  3923(198) &  0.577(58) &    0.066(10) & $11^{h}16^{m}28\fs114$ & $-03\arcdeg58\arcmin31\farcs58$ & 14.887 & 12.422 & 11.712 & 11.577 & \\
   201208431 &   K7.7(1.2) &  0.658(56) &  3900(195) &  0.678(55) &    0.090(12) & $11^{h}38^{m}58\fs954$ & $-03\arcdeg54\arcmin20\farcs11$ & 14.409 & 12.367 & 11.747 &11.571 &  \\
   201367065 &   M0.1(1.1) &  0.565(61) &  3976(205) &  0.584(63) &    0.072(12) & $11^{h}29^{m}20\fs388$ & $-01\arcdeg27\arcmin17\farcs23$ & 11.574 & 9.421 & 8.805 & 8.561 & \\
   201465501 &   M2.8(0.6) &  0.366(53) &  3460(164) &  0.369(61) &   0.0173(36) & $11^{h}45^{m}03\fs472$ & $+00\arcdeg00\arcmin19\farcs08$ &14.957 &12.451 & 11.710 & 11.495 & \\
   201617985 &   M0.5(0.7)  & 0.606(39) &  3853(135) &  0.626(39) &    0.0975(70) & $11^{h}57^{m}57\fs998$ & $+02\arcdeg19\arcmin17\farcs31$ & 14.110 & 11.719 & 11.094 & 10.900 & 8 \\
   201690311 &   K4.2(1.2) &  0.697(94) &  3948(203) &  0.714(92) &    0.106(21) & $11^{h}49^{m}16\fs849$ & $+03\arcdeg28\arcmin32\farcs05$ & 15.288 & 13.463 & 12.873 & 12.729 & \\
   201717274 &   M3.0(0.8) &  0.368(80) &  3528(165) &  0.373(92) &   0.0188(59) & $11^{h}35^{m}18\fs664$ & $+03\arcdeg56\arcmin02\farcs96$ & 14.828 & 12.911 & 12.367 & 12.153 & \\
   201912552 &   M3.0(0.9) &  0.411(53) &  3527(162) &  0.419(58) &   0.0234(44) & $11^{h}30^{m}14\fs510$ & $+07\arcdeg35\arcmin18\farcs21 $& 12.473 & 9.763 & 9.135 & 8.899 & \\
   204489514 &   M2.7(1.3) &  0.230(56) &  3096(198) &  0.207(71) &   0.0044(15) & $16^{h}03^{m}01\fs616$ & $-22\arcdeg07\arcmin52\farcs40 $& 14.080 & 12.731 & 12.110 & 11.729 & \\
   205145448 &   M1.8(1.9) &  0.402(52) &  4035(259) &  0.409(59) &   0.0384(75) & $16^{h}33^{m}47\fs672$ & $-19\arcdeg10\arcmin40\farcs04 $& 13.651 & 10.977 & 10.351 & 10.120 & \\
   205916793 &   M0.0(0.8) &  0.707(60) &  4103(230) &  0.724(56) &    0.127(17) & $22^{h}32^{m}13\fs004$ & $-17\arcdeg32\arcmin38\farcs38 $& 13.441 & 11.850 & 11.231 & 11.075 & \\
   205924614 &   K4.2(1.2) &  0.769(63) &  4240(259) &  0.785(59) &    0.172(22) & $22^{h}15^{m}00\fs462$ & $-17\arcdeg15\arcmin02\farcs55 $& 13.087 & 11.230 & 10.615 & 10.471 & 7 \\
   206011691 &   K7.9(1.1) &  0.721(59) &  3952(202) &  0.737(56) &    0.114(14) & $22^{h}41^{m}12\fs885$ & $-14\arcdeg29\arcmin20\farcs35 $& 12.316 & 10.251 & 9.633 & 9.417 & 7 \\
   206061524 &   M0.7(1.2) &  0.726(62) &  3961(213) &  0.743(59) &    0.117(15) & $22^{h}20^{m}13\fs766$ & $-13\arcdeg06\arcmin52\farcs66 $& 14.443 & 12.413 & 11.796 & 11.579 & \\
   206162305 &   M1.1(1.1) &  0.695(58) &  3896(202) &  0.713(56) &    0.100(13) & $22^{h}23^{m}02\fs289$ & $-10\arcdeg29\arcmin18\farcs89 $& 14.807 & 12.608 & 11.933 & 11.766 & \\
   206192813 &   M1.6(1.2) &  0.622(62) &  3966(225) &  0.642(62) &    0.086(13) & $22^{h}46^{m}53\fs865$ & $-09\arcdeg52\arcmin53\farcs83 $& 14.875 & 12.598 & 11.927 & 11.732 & \\
   206209135 &   M2.7(0.9) &  0.359(54) &  3370(166) &  0.361(61) &   0.0149(32) & $22^{h}18^{m}29\fs271$ & $-09\arcdeg36\arcmin44\farcs58 $& 14.407 & 11.685 & 11.122 & 10.962 & 7 \\
   211331236 &   M1.0(1.0) &  0.467(66) &  3781(203) &  0.481(71) &   0.0400(82) & $08^{h}55^{m}25\fs364$ & $+10\arcdeg28\arcmin08\farcs87 $& 13.905 & 11.447 & 10.801 & 10.589 & 7 \\
   211357309 &   M2.1(0.7) &  0.506(54) &  3790(179) &  0.523(57) &   0.0474(75) & $08^{h}52^{m}55\fs831$ & $+10\arcdeg56\arcmin41\farcs00 $& 13.155 & 10.781 & 10.165 & 9.885 & 7 \\
   211428897 &   M2.7(1.2) &  0.324(54) &  3577(200) &  0.321(63) &   0.0155(37) & $08^{h}35^{m}25\fs812$ & $+12\arcdeg04\arcmin33\farcs04 $& 13.205 & 10.414 & 9.863 & 9.624 & 7 \\
   211770795 &   K4.3(1.3) &  0.637(58) &  4311(291) &  0.656(58) &    0.126(18) & $08^{h}48^{m}02\fs336$ & $+16\arcdeg54\arcmin06\farcs67 $& 14.489 & 12.841 & 12.265 & 12.174 & 7 \\
   211799258 &   M3.0(0.8) &  0.271(54) &  3411(176) &  0.257(66) &   0.0089(26) & $08^{h}32^{m}59\fs077$ & $+17\arcdeg18\arcmin23\farcs57 $& 15.979 & 13.017 & 12.420 & 12.185 & 7 \\
   211831378 &   M0.3(1.5) &  0.608(61) &  4141(250) &  0.627(61) &    0.098(15) & $08^{h}24^{m}33\fs033$ & $+17\arcdeg45\arcmin43\farcs16 $& 16.270 & 13.972 & 13.228 & 13.085 & \\
   211916756 &   M1.2(0.9) &  0.420(90) &  3704(214) &   0.43(10) &   0.0299(93) & $08^{h}37^{m}27\fs058$ & $+18\arcdeg58\arcmin36\farcs07 $& 15.498 & 13.312 & 12.738 & 12.474 & \\
   211970234 &   M4.2(0.8) &  0.185(57) &  3000(213) &  0.148(74) &   0.0025(11) & $09^{h}04^{m}21\fs043$ & $+19\arcdeg46\arcmin48\farcs98 $& 16.122 & 13.851 & 13.186 & 12.987 & \\
   212006344 &   M0.8(0.9) &  0.595(55) &  3918(226) &  0.615(56) &    0.075(11) & $08^{h}25^{m}54\fs315$ & $+20\arcdeg21\arcmin34\farcs45 $& 12.466 & 10.104 & 9.457 & 9.275 & 7 \\
   212069861 &   M0.6(0.9) &  0.692(58) &  4078(223) &  0.709(56) &    0.119(15) & $08^{h}57^{m}46\fs605$ & $+21\arcdeg27\arcmin12\farcs72 $& 14.102 & 11.907 & 11.250 & 11.055 & 7 \\
   212154564 &   M2.7(1.2) &   0.32(15) &  3502(162) &   0.32(18) &   0.0140(94) & $08^{h}54^{m}33\fs884$ & $+23\arcdeg07\arcmin58\farcs40 $& 15.105 & 12.838 & 12.227 & 11.975 & 7 \\
   212315941 &   K7.9(0.9) &   0.48(13) &  4056(219) &   0.49(14) &    0.057(22) & $13^{h}32^{m}20\fs944$ & $-17\arcdeg03\arcmin40\farcs29 $& 14.406 & 12.844 & 12.295 & 12.175 & \\
   212354731 &   M1.8(1.1) &  0.356(55) &  3369(166) &  0.356(62) &   0.0147(33) & $13^{h}33^{m}22\fs379$ & $-16\arcdeg00\arcmin23\farcs85 $& 15.805 & 13.412 & 12.822 & 12.507 & 7 \\
   212565386 &   M1.0(0.8) &  0.570(56) &  3989(228) &  0.590(58) &    0.074(11) & $13^{h}30^{m}26\fs554$ & $-11\arcdeg20\arcmin29\farcs42 $& 14.727 & 12.368 & 11.746 & 11.513 & 7 \\
   212679798 &   M0.6(1.1) &  0.545(53) &  3716(171) &  0.563(55) &   0.0508(74) & $13^{h}29^{m}56\fs550$ & $-08\arcdeg44\arcmin58\farcs70 $& 14.846 & 13.056 & 12.469 & 12.338 & 7 \\
   212756297 &   K4.6(0.8) &  0.717(60) &  4242(257) &  0.735(58) &    0.150(20) & $13^{h}50^{m}37\fs408$ & $-06\arcdeg48\arcmin14\farcs42 $& 13.009 & 11.350 & 10.794 & 10.619 & \\
   212773309 &   M0.6(0.8) &  0.506(56) &  3886(199) &  0.523(59) &   0.0524(86) & $13^{h}49^{m}32\fs380$ & $-06\arcdeg19\arcmin21\farcs87 $& 11.391 & 9.802 & 9.272 & 9.114 & 7 \\
\enddata\label{tab:stars}
\tablenotetext{a}{Stars that are not Gliese stars (GJ) are the EPIC ID of {\it K2} stars.}
\tablenotetext{b}{Notes indicate stars with
  interferometrically determined radii and temperatures from (1)
  \cite{vonbraun:2014}, (2) \cite{boyajian:2012b}, (3)
  \cite{vonbraun:2012}, (4) \cite{demory:2009}, (5)
  \cite{vonbraun:2011}, (6) \cite{boyajian:2008}.  (7)
  indicates those stars with parameters reported in the companion
  paper by \cite{dressing:2017}. Finally, (8) indicates averages 
  of spectra obtained between two separate nights.}
\end{deluxetable}
\clearpage
\end{turnpage}

\begin{deluxetable}{llrlllllll}
\tabletypesize{\scriptsize}\tablecaption{K2 Planet and Candidate Parameters}\tablehead{
Name \tablenotemark{a}&   EPIC &        $P$ &      $R_*$ &  $T_{eff}$ &      $M_*$ &          $a$ &  $S_{inc}$ &      $R_P$ &  $T_{eq}$ \\
     &        &      [d]     &    [$R_\odot$] &  [K]  &      [$M_\odot$] &    [AU]       & [$S_\oplus$] & [$R_\oplus$] & [K]\\
}\startdata
K2-43b &  201205469.01 &   3.471140 &  0.559(57) &  3923(198) &  0.577(58) &   0.0374(13) &  47.7(8.2) &   4.01(45) &   720 \\
K2-4b  &  201208431.01 &  10.004438 &  0.658(56) &  3900(195) &  0.678(56) &   0.0800(22) &  14.0(2.0) &   2.52(37) &   530 \\
K2-3b &   201367065.01 &  10.054428 &  0.565(61) &  3976(205) &  0.584(62) &   0.0762(27) &  12.4(2.2) &   2.15(26) &   510 \\
K2-3c &   201367065.02 &  24.643479 &  0.565(61) &  3976(205) &  0.584(62) &   0.1384(50) &   3.72(68) &   1.76(22) &   380 \\
K2-3d &   201367065.03 &  44.560906 &  0.565(61) &  3976(205) &  0.584(62) &   0.2055(75) &   1.70(30) &   1.44(18) &   310 \\
K2-9b &   201465501.01 &  18.447385 &  0.366(53) &  3460(164) &  0.370(61) &   0.0980(56) &   1.78(45) &   4.9(1.1)  &   320 \\
 &  201617985.01 &   7.281384 &  0.608(55) &  3868(193) &  0.627(56) &   0.0630(19) &  19.0(2.9) &     27(23) &   570 \\
K2-49b &  201690311.01 &   2.770645 &  0.697(94) &  3948(203) &  0.713(92) &   0.0346(15) &     89(18) &   2.90(44) &   840 \\
 &  201717274.01 &   3.527432 &  0.368(80) &  3528(165) &  0.371(93) &   0.0326(29) &  18.1(6.8) &   1.55(39) &   560 \\
K2-18b &  201912552.01 &  32.941798 &  0.411(53) &  3527(162) &  0.419(58) &   0.1502(70) &   1.04(23) &   2.31(31) &   280 \\
 &  204489514.01 &  10.223626 &  0.230(56) &  3096(198) &  0.206(71) &   0.0544(66) &   1.5(1.1) &  14.8(6.6) &   300 \\
K2-54b &  205916793.01 &   9.784339 &  0.707(60) &  4103(230) &  0.725(57) &   0.0804(21) &  19.8(2.8) &   2.10(27) &   580 \\
K2-55b &  205924614.01 &   2.849258 &  0.769(63) &  4240(259) &  0.784(59) &  0.03620(90) &    131(18) &   4.63(40) &   920 \\
K2-21b &  206011691.01 &  9.323890 &  0.721(59) &  3952(202) &  0.739(57) &   0.0786(20) &  18.4(2.5) &   1.92(18) &   560 \\
K2-21c &  206011691.02 &   15.501158 &  0.721(59) &  3952(202) &  0.739(57) &   0.1101(28) &   9.4(1.3) &   2.37(24) &   480 \\
 &  206061524.01 &   5.879750 &  0.726(62) &  3961(213) &  0.742(58) &   0.0576(15) &  35.0(5.2) &   6.92(61) &   660 \\
K2-69b &  206162305.01 &   7.065991 &  0.695(58) &  3896(202) &  0.714(56) &   0.0644(17) &  24.1(3.3) &   3.25(37) &   600 \\
K2-71b &  206192813.01 &   6.985406 &  0.622(62) &  3966(225) &  0.640(62) &   0.0615(20) &  22.9(3.7) &   3.11(42) &   600 \\
K2-72b &  206209135.01 &   5.577387 &  0.359(54) &  3370(166) &  0.362(61) &   0.0438(26) &   7.8(2.0) &   1.15(20) &   460 \\
K2-72c &  206209135.02 &  15.187114 &  0.359(54) &  3370(166) &  0.362(61) &   0.0855(50) &   2.05(53) &   1.30(22) &   330 \\
K2-72d &  206209135.03 &   7.759932 &  0.359(54) &  3370(166) &  0.362(61) &   0.0548(32) &   5.0(1.3) &   1.10(21) &   410 \\
K2-72e &  206209135.04 &  24.166851 &  0.359(54) &  3370(166) &  0.362(61) &   0.1163(67) &   1.11(28) &   1.25(24) &   280 \\
 &  211331236.01 &   1.291651 &  0.467(66) &  3781(203) &  0.481(71) &  0.01819(92) &    119(28) &   1.46(57) &   900 \\
 &  211428897.01 &   1.610918 &  0.324(54) &  3577(200) &  0.320(64) &   0.0183(13) &     46(13) &   0.86(16) &   710 \\
 &  211770795.01 &   7.729341 &  0.637(58) &  4311(291) &  0.656(57) &   0.0665(19) &  28.2(4.5) &   2.3(1.3) &   630 \\
 &  211799258.01 &  19.535120 &  0.271(54) &  3411(176) &  0.257(66) &   0.0900(82) &   1.12(41) &  15.7(7.9) &   280 \\
K2-95b &  211916756.01 &  10.133866 &  0.420(90) &  3704(214) &   0.43(10) &   0.0691(60) &   6.2(2.4) &     13(14) &   430 \\
 &  211970234.01 &   1.483459 &  0.185(57) &  3000(213) &  0.150(75) &   0.0136(23) &     13(35) &  10.0(6.8) &   520 \\
 &  212006344.01 &   2.219215 &  0.595(55) &  3918(226) &  0.615(56) &  0.02835(87) &     93(14) &   1.28(15) &   850 \\
 &  212069861.01 &  30.953052 &  0.692(58) &  4078(223) &  0.709(56) &   0.1721(44) &   4.01(57) &   3.12(36) &   390 \\
 &  212154564.01 &   6.413647 &   0.32(15) &  3502(162) &   0.32(18) &   0.0464(91) &   7.1(0.3) &   2.5(1.2) &   450 \\
 &  212315941.01 &  12.935695 &   0.48(13) &  4056(219) &   0.50(14) &   0.0851(87) &   8.0(4.1) &   5.9(3.7) &   460 \\
 &  212354731.01 &  20.397357 &  0.356(55) &  3369(166) &  0.358(63) &   0.1037(60) &   1.34(36) &  25.5(9.7) &   290 \\
 &  212679798.01 &   1.834810 &  0.545(53) &  3716(171) &  0.562(55) &  0.02417(80) &     86(13) &     28(18) &   830 \\
 &  212756297.01 &   1.337116 &  0.717(60) &  4242(257) &  0.733(58) &  0.02142(56) &    326(45) &  13.7(1.2) &  1160 \\
\enddata\label{tab:candidates}
\tablenotetext{a}{{\it K2} names indicate validated planets, while those without a {\it K2} name indication remain planet candidates.}
\end{deluxetable}

\end{document}